\documentclass[12pt,preprint]{aastex}

\shorttitle{An Outflow from QU~Carinae}
\shortauthors{Kafka et al.}

\begin{document}


\title{QU~Carinae: a SNeIa progenitor?}


\author{S. Kafka\altaffilmark{1,2,3}}
\affil{Cerro Tololo Inter-American Observatory (CTIO), Casilla 603, La Serena, Chile}
\author{R. Anderson\altaffilmark{1}}
\affil{McMaster University, Department of Physics \& Astronomy ABB-241, 1280 Main St. W. Hamilton, ON L8S 4M1 Canada}

 \and 

\author{R.K. Honeycutt}
\affil{Astronomy Department, Indiana University, Swain Hall West, Bloomington, IN 47405 USA}

\altaffiltext{1}{Visiting Astronomer, Cerro Tololo Inter-American Observatory.
CTIO is operated by AURA, Inc.\ under contract to the National Science
Foundation.}
\altaffiltext{2}{Present address: Spitzer Science Center/Caltech, MS 220-6,
1200 E.California Blvd, Pasadena, CA 91125.}
\altaffiltext{3}{e-mail:stella@caltech.edu}


\begin{abstract}

Optical spectra obtained in 2006-07 of the nova-like cataclysmic variable 
QU~Car are studied for radial velocities, line profiles, and line identifications.
We are not able to confirm the reported 10.9 hr orbital period from 1982,
partly because our sampling is not ideal for this purpose and also,
we suspect, because our radial velocities are distorted by line profile
changes due to an erratic wind.  P-Cygni profiles are found in several of
the emission lines, including those of C IV.  Carbon lines are abundant in 
the spectra, suggesting a carbon enrichment in the doner star.  The presence
of [O III] 5007$\AA$ and [N II] 6584$\AA$ is likely due to a diffuse nebula
in the vicinity of the system.

The wind signatures in the spectra and the presence of nebular lines
are in agreement with the accretion wind evolution scenario that has been suggested 
to lead to SNeIa.  We argue that QU~Car is a member of the V~Sge subclass of CVs, 
and a possible SNeIa progenitor.  It is shown that the recent light curve of
QU~Car has $\sim$1 mag low states, similar to the light curve of V~Sge,
strengthening the connection of QU~Car with V~Sge stars, supersoft x-ray
sources, and SNeIa progenitors.

\end{abstract}



\section{Introduction}

Although identification of the progenitors of Supernovae Ia (SNeIa) 
remains controversial, 
it is accepted that they originate in binary systems in which at least one 
component is a white dwarf (WD); those systems are grouped under 
the wide umbrella of cataclysmic variables (CVs). CVs are semi-detached 
binaries in which a WD is accreting material from its lower main 
sequence or evolved companion (Warner 1995). When the magnetic field of the 
WD is low, a luminous accretion disk leads the gas onto the WD; if the 
magnetic field of the WD exceeds $\sim$10MG, gas is guided onto 
the WD's magnetic poles via its magnetic field lines. Depending on the
orbital period, the mass transfer rate ($\dot{M}$), the white dwarf temperature and the 
donor star's nature, CVs present a rich group of members, each providing 
valuable information on accretion physics, stellar evolution, binary star 
interaction and cataclysmic explosions. 

Current theories for SNeIa progenitors hold that, either via Roche lobe 
overflow of the companion or via a wind, the WD accumulates H or He-rich 
material which is then burned to C and O on the WD. Under suitable conditions 
(determined primarily by the stability of $\dot{M}$ onto the 
WD) the WD mass reaches the Chandrasekhar limit initiating a series of 
thermonuclear reactions eventually leading to a SNeIa (Hillebrandt \&
Niemeyer 2000). Although outlining 
this scenario is rather straightforward, the specifics are far from 
being understood or defined. For example, the nature of the companion star 
is uncertain, as well as how $\dot{M}$ remains low enough 
to prevent the expansion of the star's outer layers but high enough to allow 
stable H or He burning on the WD atmosphere without leading to a nova explosion. 
The WD mass necessary to initiate the thermonuclear reactions leading to the 
explosion is about 1.4M$_{\sun}$ and recent theoretical models conclude that a 
fast rotating WD with initial mass as low as 0.7M$_{\sun}$ can accrete efficiently 
and eventually reach the Chandrasekhar limit (Hachisu, Kato \& Nomoto 2007).  Currently, candidate 
progenitors are either double-degenerate (WD+WD) systems, or binaries that 
harbor a post main sequence companion (the single degenerate binary, or SDB, 
scenario). Double-degenerate systems are less promising 
candidates due to their low overall mass or long orbital periods (Parthasarathy 
et al. 2007). Favorable candidates are the SDBs (Langer et al. 2000) for which 
stellar evolution codes predict the presence of substantial outflows; the 
outflows result from H- or He-rich material that is processed on the WD and 
expelled under appropriate conditions (e.g. Han \& Podsiadlowski 2004). 
Badenes et al. (2007) predict that the presence of stable, high-velocity 
winds (v$_{outflow}$$\ge$1000km/sec) should lead to the formation of cavities 
in the surrounding ISM. However, such cavities are not observed around existing 
SNeIa remnants. Badenes et al. (2007) suggest that if outflows exist from SNeIa 
progenitors, they should be irregular and variable in time and should not be 
able to affect the surrounding pre-SN environment.

Among the most promising SNeIa progenitors are some types of transient 
supersoft x-ray sources (SSS), recurrent novae, and V~Sge-type CVs. 
The latter is a small category of semi-detached binaries with high mass transfer 
rates  ($\sim$10$^{-7}$M$_{\sun}$yr$^{-1}$) allowing nuclear burning on the white 
dwarf as mass transfer proceeds; these are considered to be the galactic 
counterparts of extragalactic SSS sources. Hachisu $\&$ Kato (2003; hereafter HK03) 
model the long-term optical light curve of the prototype of the class, V~Sge, 
and correlate the observed bright/faint states ($\sim$1 mag in amplitude) of the 
system with the predictions of their accretion wind evolution scenario (AWE). 
In short, the mass of the WD increases until it reaches a critical 
limit at which time the WD burns the accumulated H and expels a 
large portion of the processed material (primarily C and O) in a fast wind. This 
wind sweeps away the disk surface increasing its luminosity (leading to an overall 
increase of the system brightness by up to 1 mag) and strips off the outer layers 
of the secondary star which becomes smaller than its Roche lobe; this halts mass 
transfer. The disk is then accreted onto the WD. Mass accretion eventually stops, 
leading to a cessation of the wind phase. Mass transfer starts again with the 
donor star regaining contact with its Roche lobe; this is also the beginning of 
a new cycle. At the same time, the white dwarf accumulates enough material to 
eventually reach the Chandrasekhar limit and finally explode as a SNeIa.  
 
In this paper we present a spectroscopic study of QU~Car, an unusual yet 
bright and understudied CV. QU~Car was first reported in
Stephenson et al. (1968) as an irregular variable 
HD310376, similar to the bright low-mass X-ray binary Sco X-1 and the old nova 
HR Lyr. QU~Car shares with Sco X-1 rapid and erratic photometric variations, 
as well as marked changes in the emission line spectrum.  However, in contrast 
to Sco X-1, QU~Car sometimes has He absorption features and relatively weak x-ray 
emission. (Schild 1969). In the optical, the system is non-eclipsing and exhibits 
flickering of 
0.1-0.2 mag on timescales of 1-10 min (Schild 1969; Gilliland $\&$ Phillips 
1982, hereafter GP82).  
In our study, we present new optical spectra suggesting that QU~Car could be a 
member of the V~Sge category and a likely SNeIa progenitor. The following 
section presents our data and data reduction technique, followed by the results 
of our analysis in $\S3$ and our discussion in $\S4$. We summarize our 
findings in $\S5$.

\section{Data Acquisition and Reduction}

All our spectroscopic data were obtained with the RC Spectrograph on the
CTIO 4-m Blanco telescope during four nights in 2006 and 2007. For each run, 
we used the KPGL3 grating, yielding a spectrum from 3800-7500$\AA$ at 
1.2$\AA$/pixel, providing a resolution of $\sim$3$\AA$. A log of our observations 
is provided in Table 1.  For detector calibrations, we used the standard 
IRAF\footnote{IRAF is distributed by the National Optical
  Astronomy Observatories, which are operated by the Association of
  Universities for Research in Astronomy, Inc., under cooperative
  agreement with the National Science Foundation.} 
procedures, and for data reductions we used IRAF's onedspec/twodspec packages. 

Due to passing clouds we had strong telluric features at $\sim$5920-6020$\AA$
and 6447-6610$\AA$ for the 2006-Dec and 2007-Mar data. To correct for this 
absorption, we used data from the Wallace and Livingston (2003) atlas and the 
IRAF task ``telluric''.  Using ``telluric'' we interactively scaled the telluric 
features  to match the observed spectra in regions having few stellar features; 
the goal was to minimize the RMS difference in these regions. For all our 
spectra, we achieved a residual between the scaled correction spectrum 
and the spectrum of our star of $\lesssim$0.008 (rms) for the selected regions. 
For the nights when we had 
passing clouds no standard stars were obtained; therefore the 2006-Dec and 
2007-Mar data are not flux calibrated. For the 2007-Jun data we used 
LTT 4816 for flux calibration. The radial velocity corrected and summed 
spectrum of QU~Car appears in Figure~1 (top) with prominent 
emission/absorption features identified; summed individual spectra 
from each epoch of observations are also plotted.           

\section{Analysis}

\subsection{Radial Velocities}
We used IRAF/splot with a Gaussian fit to derive radial 
velocities (RVs) and equivalent widths (EWs) of the emission lines.  
Conspicious lines include H$\alpha$, H$\beta$, the CIII/NIII/OII blend 
(4640-4650 $\AA$), and He II 4686$\AA$.  The emission lines are rather weak
with respect to the continuum.  For example, the peak of the He II 4686 $\AA$
emission line is typically only 3-5\% above the continum. This HeII line
was used by GP82 to derive a binary orbital period of 
10.9h, placing QU~Car among the longer-period CVs. Table 2 lists our 
RV and EW measurements for this line.  Our data do not 
cover a full orbit at any epoch; however measurements are available at 
various phases of different cycles.  It was expected that our new data would be 
able to confirm the orbital period and provide a modern ephemeris, but that has
not been the case. We first used a periodogram (Horne $\&$ 
Balliunas 1986) on the GP82 He II data to confirm their period of 
0.454$\pm$0.014 days (10.9h). Figure 2 compares the periodogram of the 
radial velocities from GP82 with a periodogram over the same frequency
range for our new data, using both He II 4686$\AA$ and the central emission
peak of H$\alpha$. The location of 
the 10.9h period adopted by GP82 is marked, and it is apparent that the 
periodograms of the two data sets are inconsistent. The GP82 data 
consists of 49 spectra on 7 nights over 363 days.  Our new data
is comprised of 57 spectra on 4 nights over 202 days.  The two
data sets appear to be similar in time coverage, suggesting that the GP82 period
may be in error.  However, the GP82 data contain two
continuous sequences of 7.8h and 7.2h, while the longest sequences
in our new data are 1.5h and 1.6h; this difference appears
to be critical for this system. 

Figure 3 (top) shows the radial velocity curve of the 1979 data
from GP82, folded on the GP82 ``combined He II'' ephemeris.   
(The less extensive 1980 radial velocities in GP82 are not plotted because
we were not able to reproduce the GP82 phases for the 1980 data using their 
``combined HeII'' ephemeris.)  The bottom
panel shows our 2006-2007 He II RVs folded on the same ephemeris.  Too many
cycles have elapsed to expect the zeropoint of the GP82 ephemeris to remain valid
for our new data; however, we expected some coherence when folded on the
10.9h period.  Instead we see only monotonic or nearly monotonic sequences
which are more rapid than even the steepest part of the fitted curve in the
top panel.  The quoted 3\% uncertainly on the 10.9h period of GP82 accumulates
to many cycles even over the 8-month span of our data.  Therefore the
apparent sine-like curve between phases 0.3 and 0.7 in the bottom panel of
Figure 3 is not physical, because the data on either side of phase 0.5 are
separated by 8 months.  In fact, we carefully examined the power spectrum
and folded RV curves for our data over the period range 0.05-10 days, finding
no periodicities of consequence.

It appears that either the period of QU~Car is substantially shorter than
the GP82 result, or that the intervals of rapid RV changes seen in our
RV data are non-orbital in nature.  More extensive
measurements are needed to tell for sure, but we prefer the latter 
explanation given the current data.  The reason is that the line profiles 
in QU~Car are likely
distorted on time scales of 1-3 hours due to wind "events", similar to those
described in optical wind lines for some nova-like CVs by Kafka et al.
(2003) and Kafka \& Honeycutt (2004).  Evidence supporting this conclusion includes
1) QU~Car clearly has an erratic wind (from the P Cyg profiles),
2) the He II 4686 line can have a strong wind component
in some high inclination NL CVs (Honeycutt et al. 1986), and 3) the GP82 data 
also display occasional nearly monotonic changes
in RV over 1-3h that are considerably faster than any portion of
the fitted RV curve (one of these can be seen as the filled circles
in the top panel of Figure 3).  Such changes have little effect
on the period analysis of GP82 because the lengths of their spectral sequences
are considerably longer.  This is not the case with our new data, making 
it impossible to improve or confirm the GP82 period
study.  A new comprehensive RV study of QU~Car would certainly be helpful
considering the importance (and brightness) of the system.

We also measured the RVs of the other emission lines; 
however none indicated any periodicity nor did they showed any coherent 
variation with the GP82 orbital period. Schild (1969) and GP82 examined 
their short-term photometric data for periodic modulation in the V-band light curve, 
but found none. The $\sim$0.2 mag ``flaring'' events appeared to be erratic on 
timescales of a few minutes. We examined the AAVSO light curve of the system 
(Figure~4) for periodicities but were unable to find any, even when we group 
the data to reduce the scatter.

\subsection{The outflow from QU~Car and nebula lines}

In high $\dot{M}$ disk CVs, winds often produce blueshifted absorption components 
to the emission lines.  These P-Cygni profiles, most obvious in UV resonance 
lines such as C IV 1449$\AA$ and Si IV 1397$\AA$, are seen at times in both 
nova-like CVs and in dwarf novae during outburst. P-Cygni profiles appear 
mostly in low-inclination (i.e. disk almost face-on) CVs, indicating a bi-polar 
nature of the outflow. The lines are thought to form by resonance scattering in 
an accelerating wind, with velocities reaching 5000 km/s. Although the origin 
of the wind is uncertain, it is generally thought to arise from the inner
accretion disk/boundary 
layer.  Some CVs are known to have reliable P-Cygni profiles in 
optical He I and/or H lines; e.g. BZ~Cam (Ringwald $\&$ Naylor 1998) and 
Q~Cyg (Kafka et al. 2003). Those P Cygni profiles are most conspicuous in the 
He I triplet lines at 5876$\AA$ and 7065$\AA$ whereas they are absent in the He I 
singlet line at 6678$\AA$ (despite the fact that these three lines have very 
similar excitation levels). Kafka et al. (2003) argued that this behavior is 
due to the strongly metastable 2$^{3}$S effective ground state of the triplets, 
which becomes over-populated in conditions of low density and a dilute radiation 
field, over-populating in turn the 2$^{3}$P common lower level of HeI 5876$\AA$ 
and 7065$\AA$, leading to absorption in a low density wind. 

In QU~Car, variable and erratic P-Cygni profiles in N V 1238.8$\AA$/1242.8$\AA$ 
(doublet), O V 1371$\AA$ and Si IV 1398$\AA$ and C IV 1549$\AA$ have already been 
reported 
(Knigge et al. 1994; Hartley et al. 2002; Drew et al. 2003). Their 
UV P-Cygni profiles have no apparent phase dependence and unknown time evolution. 
The modest velocities of the outflow (up to $\sim$2000 km/sec) suggested an 
outer-disk origin for the wind and an unusually high $\dot{M}$ for the 
system (Hartley et al. 2002). In our (and older) optical data the 
He I triplet lines in QU~Car did not exhibit P-Cygni profiles 
(see Figure~1).  It is likely that strong C IV emission at 
5802$\AA$ overlaps any blueshifted absorption in the 
He I 5876$\AA$ line. Occasional shallow absorption dips appear in the blue side 
of the H$\alpha$ line, reaching $\sim$1000 km/sec. At the same time a broad 
P-Cygni profile is present at the blue end of the C IV 5807$\AA$ line.  
The presence of diffuse interstellar bands at 5780$\AA$, 5705$\AA$ and 6494$\AA$ 
complicate measuring the strength of this blue-shifted absorption.  However,  
one can discern {\em relative} changes in the strength of the absorption profiles 
between epochs. The EW of the absorption component of 
the CIV line is included in Table 3, indicating up to 50$\%$ 
variability in the strength of the wind at different epochs. This CIV line 
likely originates from a C V$\rightarrow$C IV recombination cascade  
requiring 490ev, indicating the presence of a strong ionization field in the 
system.  Such high ionization levels might also  
explain the lack of P-Cygni profiles in the lower-ionization He I lines.  The  
velocity of the wind in C IV reaches $\sim$5700 km/sec; this is a lower limit since 
the C III~5696$\AA$ line overlaps the P-Cygni profile at its blue end. 

We were surprised to find traces of [O III] in the spectrum of QU~Car. Of the 
two components of the doublet, only the 
5007$\AA$ component is present.   This forbidden 
line appears in planetary nebula and active galactic nuclei (AGN) as a 
result of photoionized gas around a central source (e.g. Dimitrijevic et al. 2007). 
For AGNs the flux ratio in the doublet is near 3 
(Dimitrijevic et al. 2007). If the process governing [OIII] emission in QU~Car 
is similar to that in AGNs, then 
the [OIII] 4959$\AA$ component is likely absent in QU~Car due to 
its relative faintness. The detection of forbidden lines 
indicates the presence of a nebula.   The line is weak and somewhat difficult
to measure, but it appears that the [OIII] line in QU~Car has substantial 
radial velocity variations, from $\sim$250 (epoch 3) to $\sim$-128 km s$^{-1}$ 
(epoch 2).  

In our QU~Car spectra the [O III] 5007$\AA$ line has two 
components with nearly constant displacement from the line center of 
-500 and 370 km s$^{-1}$. These likely
arise from the front and back sides of an expanding shell or wind.  We measured
the EWs of the two components using Gaussian fits ; although the red component 
hardly changes, the blue component appears to vary in strength (see Table 3).
This line is weak, but we nevertheless judge these changes to be real.
Similar behavior of [O III] 5007$\AA$ in seen in some symbiotic stars, with
the line being split into two components, and variability in strength over 
intervals of months (Kenyon 1986).  
 
The presence in QU~Car of [N II] 6584$\AA$ emission, which is also common in
planetary nebulae, confirms the existence of a nebula in the vicinity 
of the star.  Our data show EW variations in this line 
(more than 2X variation between epoch 1 and epoch 4), but again the line
is weak.  Undetected is
[Si II] (at 6717$\AA$ and 6731$\AA$), which is characteristic of the nebulosity 
associated with the strong-wind CV BZ Cam 
(Greiner et al. 2001), and visual inspection of the spectra did not indicate  
any extended nebulosity.  Overall, the presence of forbidden
lines in QU~Car is consistent with the outflows associated with the AWE
mechanism, which we argue is operating in this system to produce the
high state/low-state features in the light curve (see $\S4$), and the wind 
features in the spectrum. 

\subsection{Hydrogen and Helium emission lines}

Figure 5 demonstrates graphically the variability in the strength and profile 
of hydrogen and helium lines in QU~Car.  Tables 2 and 3 also document the
changes in EW.  In the spectroscopic study of GP82, 
the H$\beta$ emission line seems to be transient, disappearing between phases 
0.2-0.7, with a maximum EW of $\sim$3$\AA$. When present, the emission component 
of the line is superposed on a wide absorption trough. The phasing of the
RV variations of the H$\beta$ emission trailed the HeII RV curve by 
P/4.  Drew et al. (2003) searched for absorption features in 
the optical that could reveal the nature of the donor star. They discuss the 
presence of the He II~5411$\AA$ line and the complete absence of O VI 5290$\AA$, 
both of which are characteristic of SSSs.  Both GP82 and 
Drew et al. (2003) measure the He II 4686$\AA$/H$\beta$ 
EW ratio to be greater than 2, in agreement with the typical value in 
SSSs. In our data, this ratio is always less than 1.  Furthermore, in our data,
the H$\beta$ line has the largest EW values and 
EW variations ever reported in QU~Car: the EW varies from 14$\AA$ to 0.6$\AA$ 
at two successive epochs.  (Table 3). At the same time, the surrounding 
Balmer lines do not follow this variation.   We remain puzzled as to
how this situation can occur.  
The H$\alpha$ line is mostly single-peaked; however at times a second peak 
appears to be present (Figure 5). The He I 6678$\AA$, 5876$\AA$ and 7065$\AA$ 
lines seem to also have an additional (red) component at times. 

Drew et al. (2003) described an overabundance of carbon in QU~Car with respect 
to both helium and oxygen, suggesting that it is likely due to an R-type carbon 
donor star.  A number of C emission lines are identified in Figure~1; however
there is no evidence of any absorption features from the secondary star.
Perhaps the ultra-luminous accretion disk of QU~Car masks detection of the
secondary.  However, the presence 
of both C III~5696$\AA$ and C IV~5802,5812$\AA$ is typical of 
carbon Wolf-Rayet (WR) stars and carbon-type central remnants of planetary 
nebulae, suggesting that the donor star may be carbon-rich.  
Considering a wide range of possibilities, the strong carbon features 
could be due to 1) carbon
generated by nuclear burning on the white dwarf as part of the AWE cycle,
2) carbon rich gas accreted onto the WD from a carbon rich companion, or
3) a carbon WR secondary star contributing directly to the optical light
in QU~Car.  If the latter is the case, then (following the 
Crowther et al. (1998) classification criteria for WR/WC stars),
the ratio of C III~5696$\AA$ to C IV~5802,5812$\AA$ implies 
WC7 for the QU~Car secondary star.  This assumes that C III~5696$\AA$ 
is not greatly affected by the absorption component of C IV.

\section{A V~Sge star and SNeIa progenitor?}

Drew et al. (2003) used interstellar absorption features to derive 
a lower distance limit of $\sim$2kpc to QU~Car, leading to   
$\dot{M}\sim$10$^{-7}$M$_{\sun}$y$^{-1}$, consistent with high   
$\dot{M}$ in V~Sge stars. The members of this class are considered 
to have evolved donor stars that transfer gas with 
$\dot{M}\sim$10$^{-7}$M$_{\sun}$y$^{-1}$. Furthermore, synthetic UV spectra and an empirical
mass-period relation lead to a white dwarf mass of 1.2M$_{sun}$ (Linnell et al. 2008). The spectrum of QU~Car seems
similar to that of WX Cen, a member of the V~Sge category and also
a suggested SNeIA progenitor (Oliveira \& Steiner 2004).  
Overall, QU~Car appears to possess
the key properties for being a SNeIa progenitor candidate under the AWE
mechanism: very high $\dot{M}$, erratic winds (plus nebular lines), and high/low
states (as described below).  Figure 4
shows the long term AAVSO\footnote{The data presented here are from 
  the database of the American Association of Variable Star Observers (AAVSO), 
  at http://www.aavso.org/ } 
light curve of QU~Car. Beginning 
in 2002 QU~Car displayed $\sim$1 mag faint states resembling the faint states 
of the prototype of the class, V~Sge (Robertson et al. 1997).  
In V~Sge this repeated high/low
state cycle has been interpreted as evidence for the accretion wind
evolution scenario, leading to SNeIa (HK03). 
The AAVSO light curve for QU~Car shows 
that such behavior is new for this system. We may be witnessing the 
onset of a V~Sge (or SSS) phase in a CV.

\section{Summary}

We have presented a new spectroscopic study of the bright nova-like QU~Car. 
The highlights of this work are:

\begin{itemize}

\item We explored the orbital period of the system using He II radial velocities
from this study and from GP82.  We are unable to confirm the 10.9 hr period
of GP82.  We argue that this situation is lilely due to rapid variations 
(compared to orbital) in 
the line profiles (P-Cygni) of the He II line due to an erratic wind, combined
with our relatively short time series.  We use as support for this argument
the fact that the wind variability time scales in NL CVs are in this range,
and the presence of fast monotonic RV changes in our data as well as in GP82.

\item Forbidden lines are detected in the QU~Car spectrum, consistent
with substantial mass outflow.  The [O III] 5007 appears split, likely due to
contributions from the front and back sides of an expanding flow.

\item It is pointed out that the light curve of QU~Car has low states very similar
to those in V~Sge.  In V~Sge these low states are taken as evidence for the
operation of accretion wind evolution cycle that is thought to lead to
SNeIa.  We propose that QU~Car is also a strong candidate for a
SNeIa progenitor, based on this and other similarites to V~Sge systems.
These similarites include high $\dot{M}$ and evidence for a strong wind from
P Cygni profiles and nebular lines. 

\end{itemize}

It is surprising that so little optical work has been done on this
bright, important system.  It is important to determine/refine 
the orbital period of QU~Car.  Photometric monitoring is needed
to define the characteristics of the photometric cycle and allow 
accurate comparisons with the predictions of the accretion wind model. Higher 
resolution spectral data will provide information on the various ionization regions 
and possible emission lines from the nebula.  High resolution corongraphic
imaging should be employed to search for nebulosity.  
QU~Car appears to be the brightest known member of the V~Sge category, 
and a  likely SNeIa progenitor.  Further study can provide important 
information on the nature and the conditions that lead to mass accumulation 
onto the white dwarf during the wind accretion cycle.


\acknowledgments
We gratefully acknowledge observations from the AAVSO International Database 
contributed by observers worldwide and used in this research. We would also like to thank our anonymous referee for a careful review of the manuscript.

\begin{deluxetable}{cccc}
\tablenum{1}
\tablecolumns{4}
\tablecaption{Observing Log for QU~Car}
\tablehead{\colhead{UT Date}  & \colhead{$\#$ Spectra} & \colhead{Exp. time (s)} 
& \colhead{Notes}}
\startdata
 2006-Dec-12 & 19 & 180 &   partly cloudy    \\
 2007-Mar-19 & 8  & 60  &   clear; dark      \\   
 2007-Jul-01 & 2  & 180 &   clear; full moon \\
             & 21 & 240 &                    \\  
 2007-Jul-02 & 8  & 300 &                    \\
\enddata
\end{deluxetable}


\begin{deluxetable}{rrc}
\tablenum{2}
\tablecolumns{3}
\tablecaption{Radial velocities and equivalent widths of the HeII 4686$\AA$ line}
\tablehead{\colhead{HJD}  &  \colhead{RV (km s$^{-1}$)} & \colhead{EW (\AA)} }
\startdata
\hline
2454081.809  &  -136  &  -1.9  \\
2454081.812  &  -115  &  -2.0  \\
2454081.815  &  -151  &  -2.0  \\
2454081.819  &  -156  &  -2.2  \\
2454081.822  &  -182  &  -1.9  \\
2454081.825  &  -199  &  -1.8  \\
2454081.829  &  -203  &  -1.8  \\
2454081.833  &  -223  &  -1.9  \\
2454081.836  &  -217  &  -1.9  \\
2454081.839  &  -216  &  -2.0  \\
2454081.842  &  -217  &  -2.1  \\
2454081.845  &  -226  &  -2.1  \\
2454081.848  &  -224  &  -2.1  \\
2454081.852  &  -235  &  -2.3  \\
2454081.855  &  -222  &  -2.3  \\
2454081.858  &  -218  &  -2.3  \\
2454081.862  &  -197  &  -2.5  \\
2454081.865  &  -176  &  -2.5  \\
2454081.870  &  -235  &  -2.4  \\
2454178.622  &  -288  &  -1.5  \\
2454178.624  &  -266  &  -1.5  \\
2454178.627  &  -226  &  -1.5  \\
2454178.629  &  -225  &  -1.4  \\
2454178.631  &  -237  &  -1.9  \\
2454178.633  &  -257  &  -1.8  \\
2454178.635  &  -256  &  -1.6  \\
2454178.637  &  -282  &  -1.8  \\
2454282.285  &   -96  &  -2.6  \\
2454282.289  &  -111  &  -2.6  \\
2454282.293  &   -93  &  -2.6  \\
2454282.297  &   -43  &  -2.8  \\
2454282.301  &    -9  &  -2.9  \\
2454282.305  &   -14  &  -2.8  \\
2454282.309  &   -20  &  -2.8  \\
2454282.313  &   -32  &  -2.7  \\
2454283.471  &   -55  &  -2.5  \\
2454283.474  &   -68  &  -2.6  \\
2454283.477  &   -75  &  -2.5  \\
2454283.480  &   -36  &  -2.5  \\
2454283.483  &    12  &  -2.5  \\
2454283.489  &    -7  &  -2.4  \\
2454283.492  &   -23  &  -2.6  \\
2454283.495  &   -26  &  -2.5  \\
2454283.498  &   -38  &  -2.7  \\
2454283.501  &   -41  &  -2.3  \\
2454283.504  &   -35  &  -2.6  \\
2454283.507  &   -73  &  -2.6  \\
2454283.510  &   -93  &  -2.5  \\
2454283.514  &  -100  &  -2.5  \\
2454283.517  &  -106  &  -2.9  \\
2454283.520  &  -107  &  -3.0  \\
2454283.523  &  -128  &  -2.7  \\
2454283.526  &  -129  &  -3.0  \\
2454283.529  &  -120  &  -2.8  \\
2454283.532  &  -129  &  -2.7  \\
2454283.535  &  -134  &  -2.9  \\
2454283.538  &  -177  &  -2.9  \\
\enddata
\end{deluxetable}


\begin{deluxetable}{cccccc}
\tablenum{3}
\tablecolumns{6}
\tablecaption{Equivalent widths\tablenotemark{a} of the emission lines}
\tablehead{ \colhead{line\tablenotemark{b}}  &  \colhead{central wavelength} 
 &  \colhead{2006-12-12}  &  \colhead{2007-03-19}  &  \colhead{2007-07-01} 
  &  \colhead{2007-07-02}}
\startdata
\hline
H$\delta$          & 4101      &  -0.42 &  \nodata & -3.18  &  -1.54  \\
He II              & 4199      &  -0.37 &  \nodata & -0.15  &  -0.25  \\
Si II              & 4265      &  -0.39 &  -0.46   & -0.32  &  -0.26  \\
H$\gamma$          & 4340      &  -1.58 &  -0.20   & -2.11  &  -0.24  \\
He II              & 4542      &  -0.12 &  -0.22   & -0.32  &  -0.26  \\
C III/N III/O II   & 4642      &  -3.09 &  -3.08   & -3.05  &  -3.56  \\
He II              & 4686      &  -2.00 &  -1.59   & -2.75  &  -2.79  \\
H$\beta$           & 4861      & -13.94 &  -0.58   & -3.17  &  -7.60  \\
He I               & 4922      &  -0.21 &  -0.19   & -0.29  &  -0.23  \\
$[$O III$]$ (full) & 5007      &  -0.48 &  -0.46   & -0.48  &  -0.51  \\
$[$O III$]$ (blue) &           &  -0.31 &  -0.19   & -1.16  &  -0.45  \\
$[$O III$]$ (red)  &           &  -0.28 &  -0.30   & -0.24  &  -0.28  \\
He II              & 5409      &  -0.31 &  \nodata & -0.50  &  -0.45  \\
C III              & 5696      &  -0.60 &  -0.32   & -0.31  &  -0.21  \\
C IV               & 5801,5812 &  -1.17 &  -1.38   & -0.64  &  -0.92  \\
C IV abs\tablenotemark{c}   &  &   1.95 &   2.78   &  1.78  &   2.20  \\
He I               & 5875      &  -0.12 &  -0.13   & -0.31  &  -0.20  \\
H$\alpha$          & 6563      &  -3.78 &  -2.26   & -4.23  &  -3.29  \\
$[$N II$]$         & 6584      &  -0.71 &  -0.57   & -0.32  &  -0.27  \\
He I               & 6678      &  -0.67 &  -0.58   & -0.85  &  -0.79  \\
He I               & 7065      &  -0.33 &  -0.17   & -0.51  &  -0.51  \\
\enddata
\tablenotetext{a}{We follow the convention that the EW of {\em emission} 
lines are negative.}
\tablenotetext{b}{Only the EW of the Balmer line emission cores are measured}
\tablenotetext{c}{These measurements are uncertain, since 
they overlap with the DIBs at 5780$\AA$ and 5705$\AA$. Their inclusion here is to 
demonstrate the variation of the wind absorption between epochs of observations. }
\end{deluxetable}



\begin{figure}  
\epsscale{1.0}
\plotone{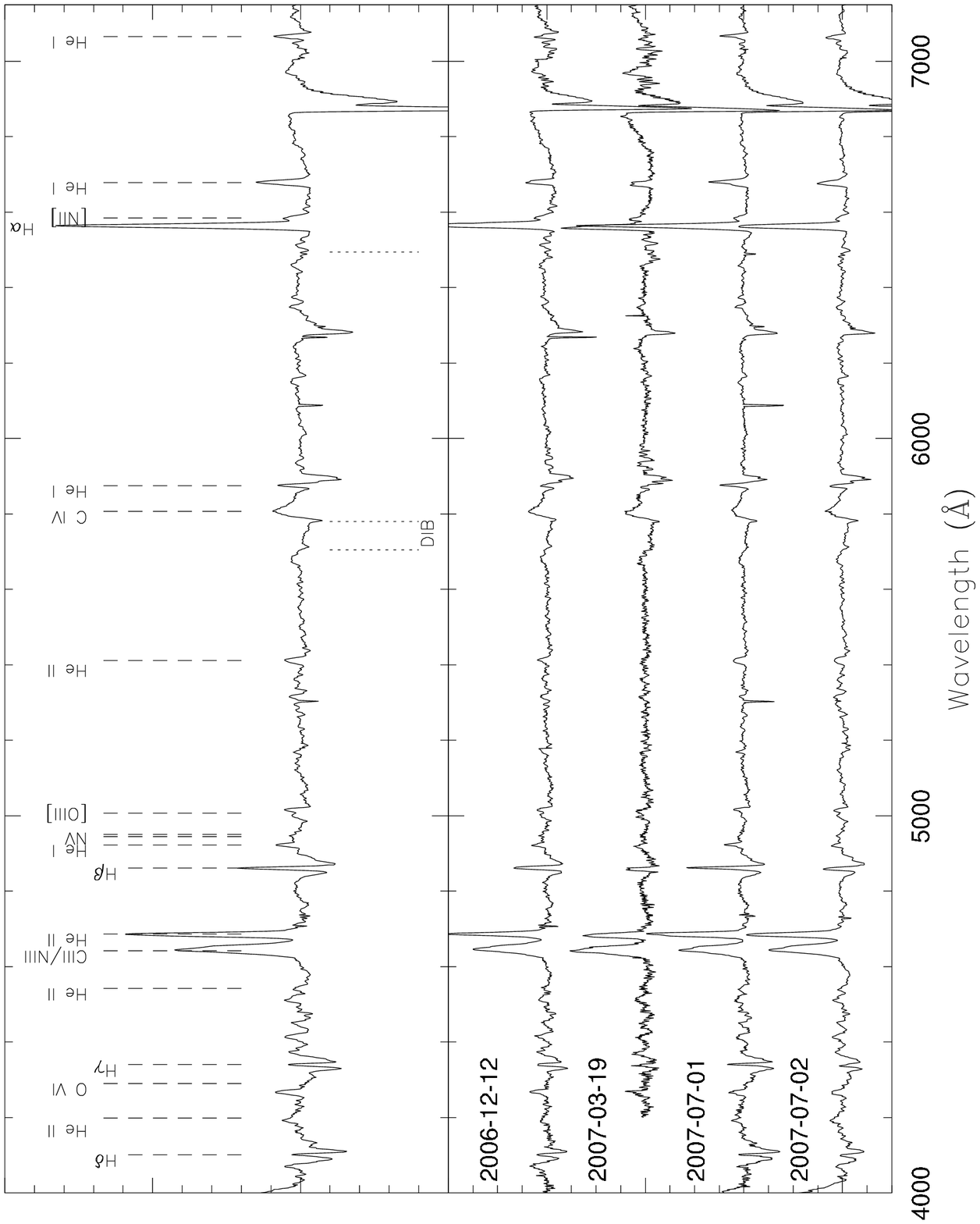}
\caption{Examples of QU~Car spectra from four epochs. The UT dates of observations 
are marked.  The top spectrum is labeled with line identifications.}
\label{example}
\end{figure}


\begin{figure}  
\epsscale{1.0}
\plotone{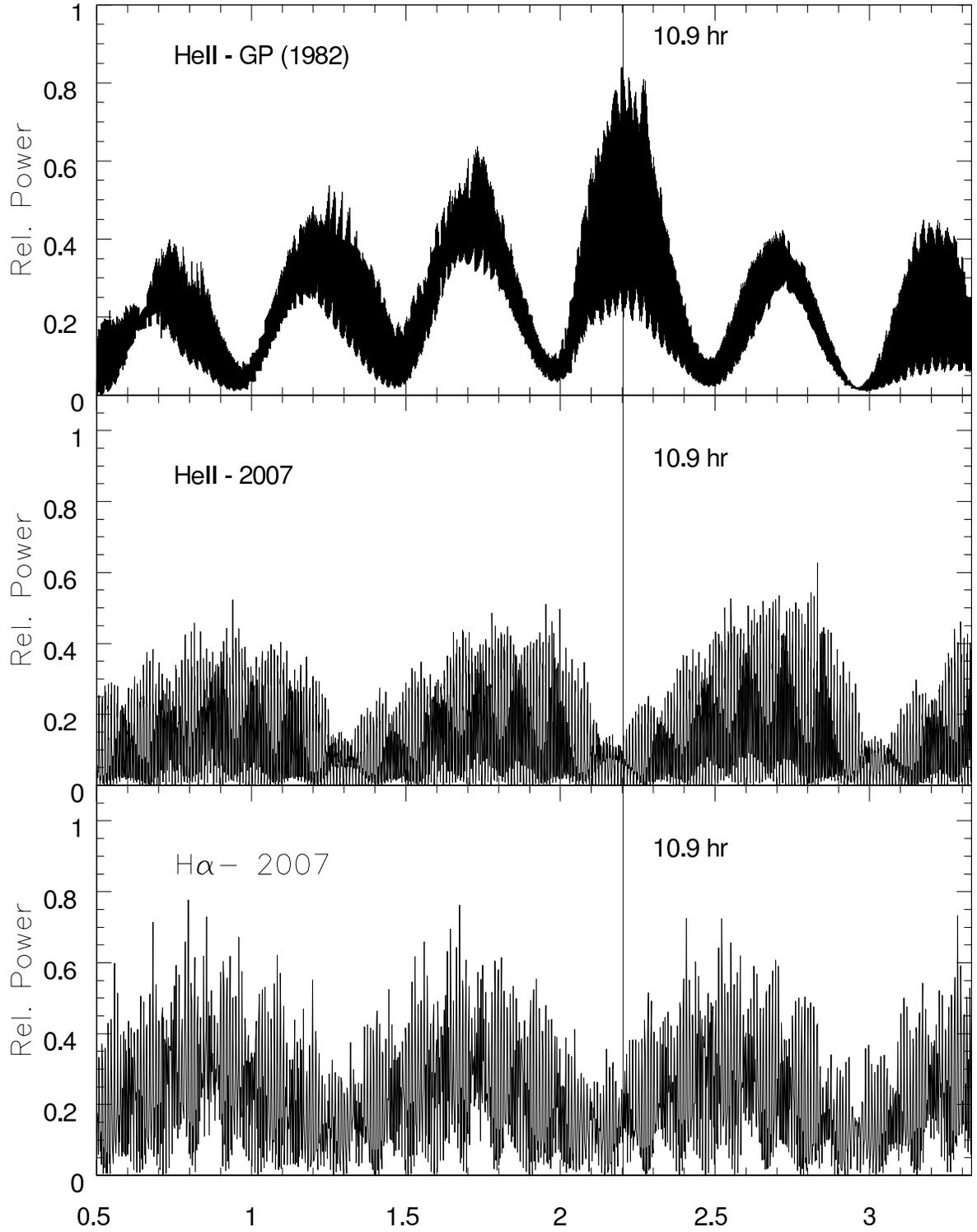}
\caption{Periodogram of the GP82 He II data (top), our He II data (middle), 
and our H$\alpha$ data (bottom). The horizontal axis is frequency in 
days$^{-1}$.  The orbital period of 10.9h from GP82 is marked in all plots.}
\label{periodogram}
\end{figure}


\begin{figure}  
\epsscale{1.0}
\plotone{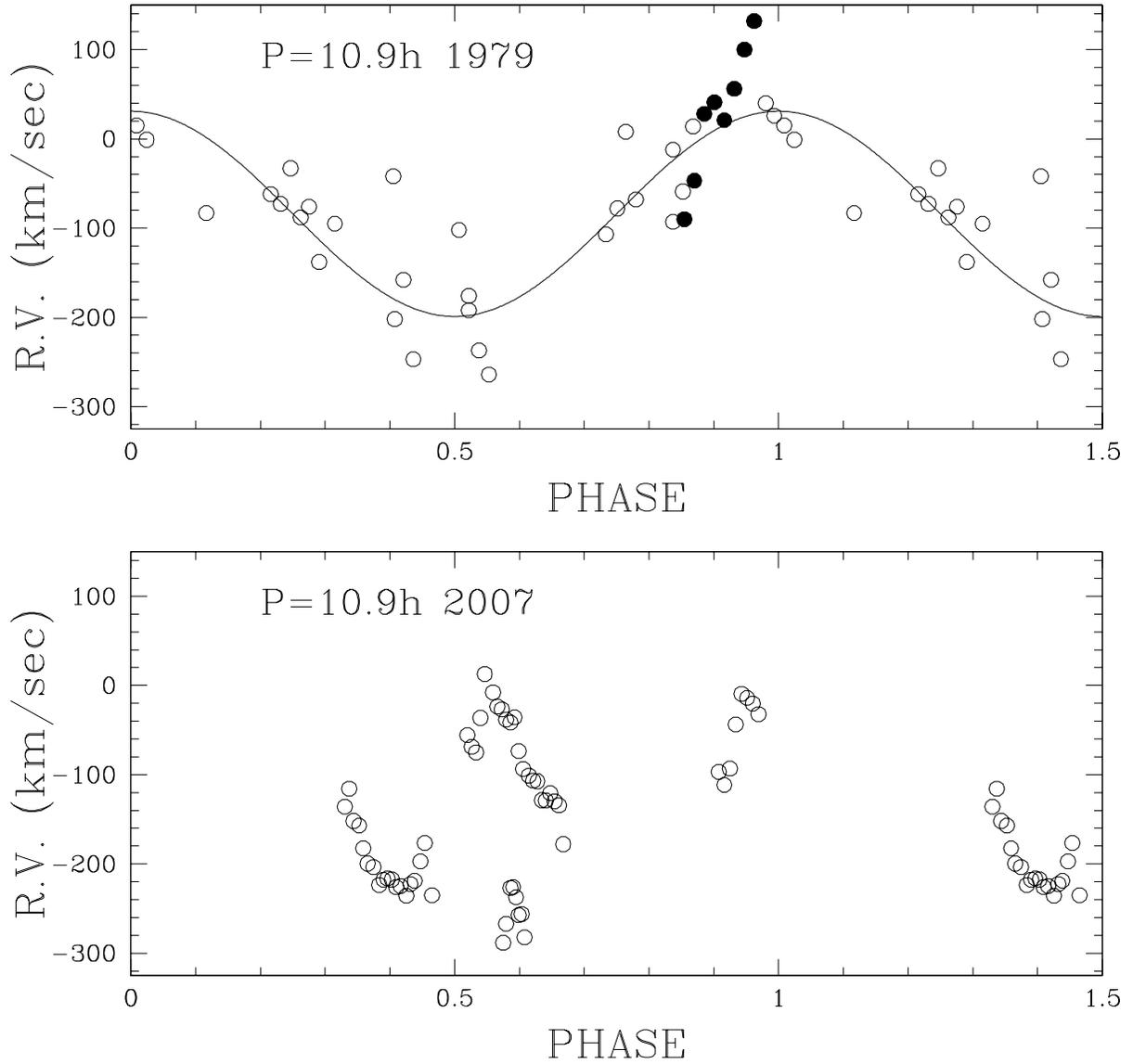}
\caption{Folded radial velocity curves of the He II 4686 line in QU~Car.  Top:
1979 data from GP82, along with their fitted curve.  The solid points mark 
a continuous sequence in time having a rapid
nearly monotonic radial velocity change.  Bottom:  2006-2007 data from
this study, showing numerous rapid monotonic changes similar to the
sequence marked in the top panel.  See text for details.}
\label{folded}
\end{figure}

\begin{figure}  
\epsscale{1.0}
\plotone{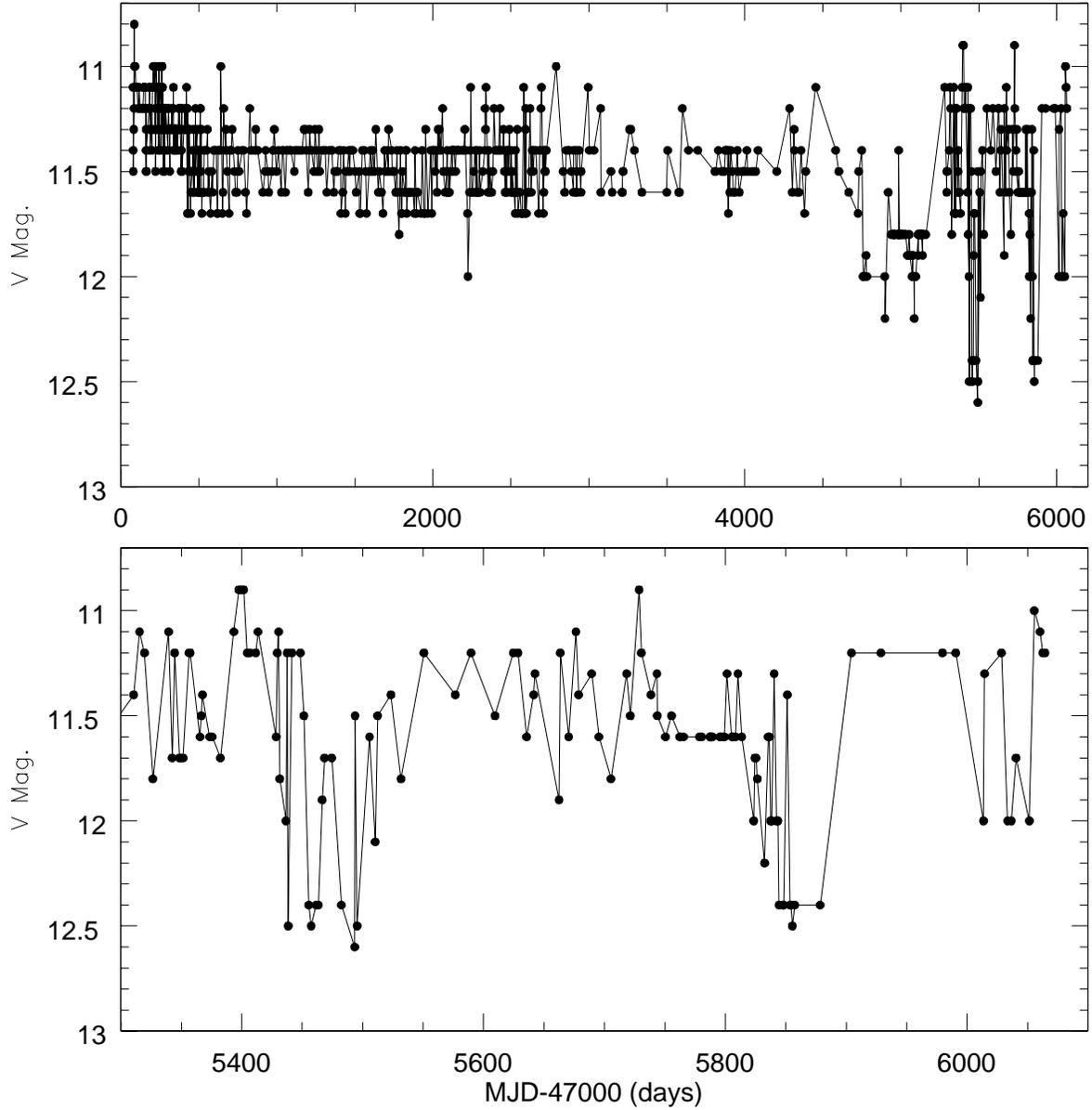}
\caption{Top: 20 years of AAVSO light curve of QU~Car. The system started showing 
V~Sge-like bright/faint states (bottom) in mid-2002 (MJD$\sim$52430).}
\label{lc}
\end{figure}

\begin{figure}  
\epsscale{0.8}
\plotone{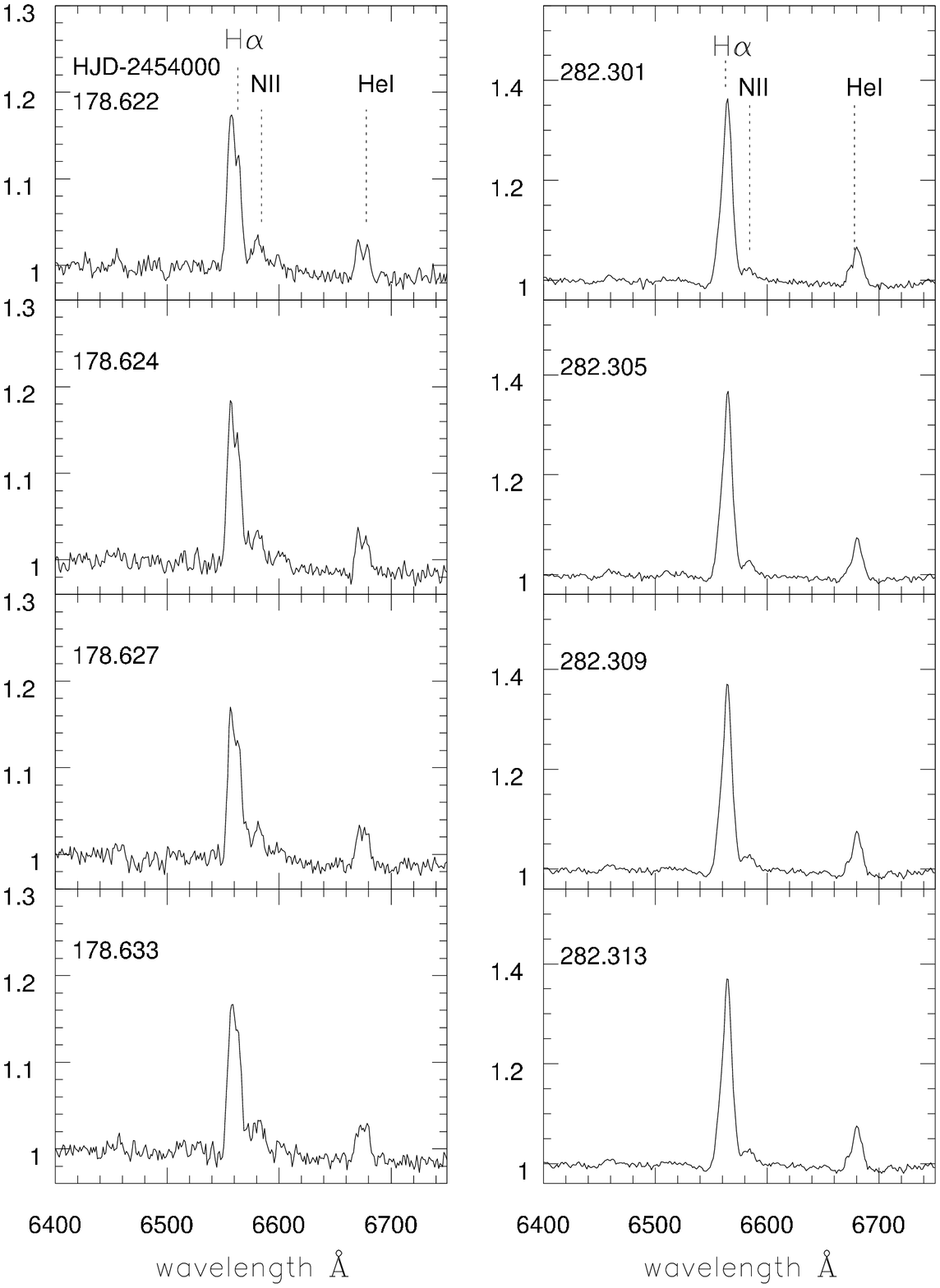}
\caption{Examples of changes in the strength and profile of H$\alpha$ and He I. 
The vertical scale is in units of the continuum and 
HJD-2454000 is shown at the top of each plot. Note the structure in both 
H$\alpha$ and He I on the left set of panels, and the lack of structure
3 months later (right panel).  Note also that the lines in 
the right  panels are twice the EW of those on the left. Finally, there is a 
weak P-Cygni profiles at times when the lines are strong (right panels).}
\label{example}
\end{figure}


\begin{references}

\reference {} Badenes, C., Hughes, 
J.~P., Bravo, E., \& Langer, N.\ 2007, \apj, 662, 472 

\reference {} Crowther, P.~A., De 
Marco, O., \& Barlow, M.~J.\ 1998, \mnras, 296, 367 

\reference {} Dimitrijevi{\'c}, M.~S., Popovi{\'c}, L.~{\v C}., Kova{\v c}evi{\'c}, J., 
Da{\v c}i{\'c}, M., \& Ili{\'c}, D.\ 2007, \mnras, 374, 1181 

\reference {} Drew, J.~E., Hartley, 
L.~E., Long, K.~S., \& van der Walt, J.\ 2003, \mnras, 338, 401 

\reference {} Gilliland, R.~L., \& Phillips, M.~M.\ 1982, (GP82) \apj, 261, 617 

\reference {} Greiner, J., et al.\ 2001, \aap, 376, 1031 

\reference {} Hachisu, I., \& Kato, M.\ 2003, (HK03) \apj, 598, 527
 
Hachisu, I., Kato, M., \& Nomoto, K.\ 2007, ArXiv e-prints, 710, arXiv:0710.0319 


\reference {} Han, Z., \& Podsiadlowski, P.\ 2004, \mnras, 350, 1301

\reference {} Hartley, L.~E., Drew, J.~E., \& Long, K.~S.\ 2002, \mnras, 336, 808 

\reference {} Hillebrandt, W., \& Niemeyer, J.~C.\ 2000, \araa, 38, 191 

\reference {} Horne, J.~H., \& Baliunas, S.~L.\ 1986, \apj, 302, 757 

\reference {} Honeycutt, R.~K., Schlegel, E.~M., \& Kaitchuck, R.~H.\ 1986, 
\apj, 302, 388 

\reference {} Kafka, S., Tappert, C., Honeycutt, R.~K., \& Bianchini, A.\ 2003, 
\aj, 126, 1472

\reference {} Kafka, S., \& Honeycutt, R.~K.\ 2004, \aj, 128, 2420 

\reference {} Kenyon, S.~J.\ 1986, The Symbiotic Stars, Cambridge and New York, 
Cambridge University Press, 1986  

\reference {} Knigge, C., Drew, J.~E., Hoare, M.~G., \& La Dous, C.\ 1994, \
mnras, 269, 891 

\reference {} Langer, N., Deutschmann,
A., Wellstein, S., Houmlflich, P.\ 2000, \aap, 362, 1046

Linnell, A.~P., Godon, P., Hubeny, I., Sion, E.~M., Szkody, P., \& Barrett, P.~E.\ 2008, ArXiv e-prints, 801, arXiv:0801.0704 

\reference {} Oliveira, A.~S., \& Steiner, J.~E.\ 2004, \mnras, 351, 685 

\reference {} Parthasarathy,
M., Branch, D., Jeffery, D.~J., \& Baron, E.\ 2007, New Astronomy Review,
51, 524

\reference {} Ringwald, F.~A., \& Naylor, T.\ 1998, \aj, 115, 286 

\reference {} Robertson, J.~W., 
Honeycutt, R.~K., \& Pier, J.~R.\ 1997, \aj, 113, 787 

\reference {} Schild, R.~E.\ 1969, \apj, 157, 709 

\reference {} Stephenson, C. B.; Sanduleak, N.; Schild, R. E. Astrophysical 
Letters, Vol. 1, p.247, 1968

\reference {} Wallace, L., \& Livingston, W.\ 2003, An atlas of the solar 
spectrum in the infrared from
1850 to 9000 cm-1 (1.1 to 5.4 micrometer), revised.~By L.~Wallace and
W.~Livingston.~ Tucson: National Solar Observatory, National Optical
Astronomy Observatory, NSO Technical Report, 2003.,

\reference {} Warner, B.\ 1995, Cataclysmic Variable Stars, Cambridge 
Astrophysics Series, Cambridge University Press  

\end{references}
\end{document}